\input harvmac
\input epsf
\noblackbox

%%% Paragraphs
\newcount\figno

\figno=0
\def\fig#1#2#3{
\par\begingroup\parindent=0pt\leftskip=1cm\rightskip=1cm\parindent=0pt
\baselineskip=11pt \global\advance\figno by 1 \midinsert
\epsfxsize=#3 \centerline{\epsfbox{#2}} \vskip 12pt
\centerline{{\bf Figure \the\figno :}{\it ~~ #1}}\par
\endinsert\endgroup\par}
\def\figlabel#1{\xdef#1{\the\figno}}
\def\pano{\par\noindent}

%%% special math symbols
\font\cmss=cmss10
\font\cmsss=cmss10 at 7pt

\def\rlx{\relax\leavevmode}
\def\inbar{\vrule height1.5ex width.4pt depth0pt}
\def\IC{\relax\,\hbox{$\inbar\kern-.3em{\rm C}$}}
\def\IR{\relax{\rm I\kern-.18em R}}
\def\IN{\relax{\rm I\kern-.18em N}}
\def\IP{\relax{\rm I\kern-.18em P}}
\def\ZZ{\rlx\leavevmode\ifmmode\mathchoice{\hbox{\cmss Z\kern-.4em Z}}
 {\hbox{\cmss Z\kern-.4em Z}}{\lower.9pt\hbox{\cmsss Z\kern-.36em Z}}
 {\lower1.2pt\hbox{\cmsss Z\kern-.36em Z}}\else{\cmss Z\kern-.4em Z}\fi}

%%% misc.
\def\narrowplus{\kern -.04truein + \kern -.03truein}
\def\narrowminus{- \kern -.04truein}
\def\narrowminussub{\kern -.02truein - \kern -.01truein}

\def\o#1{\overline{#1}}

\def\doubref#1#2{\refs{{#1},{#2}}}
\def\threeref#1#2#3{\refs{{#1},{#2},{#3} }}

%%% further macros

\def\br{\hfill\break}
\def\h {{1\over 2}}
\def\fc#1#2{{#1 \over #2}}
\def\lf {\left}
\def\ri {\right}

%%% References

%\AngelantonjCT
\lref\rasreview{
C.~Angelantonj and A.~Sagnotti,
{\it Open Strings},
hep-th/0204089.
%%CITATION = HEP-TH 0204089;%%
}

%\GinspargEE
\lref\GinspargEE{
P.~Ginsparg,
{\it Gauge And Gravitational Couplings In Four-Dimensional String Theories},
Phys.\ Lett.\ B {\bf 197}, 139 (1987).
%%CITATION = PHLTA,B197,139;%%
}

%\IbanezBD
\lref\IbanezBD{
L.E.~Ibanez,
{\it Gauge coupling unification: Strings versus SUSY GUTs},
Phys.\ Lett.\ B {\bf 318} (1993) 73, hep-ph/9308365.
%%CITATION = HEP-PH 9308365;%%
}

%\EllisZQ
\lref\EllisZQ{
J.R.~Ellis, S.~Kelley and D.V.~Nanopoulos,
{\it Precision Lep Data, Supersymmetric Guts And String Unification},
Phys.\ Lett.\ B {\bf 249}, 441 (1990);\br
%%CITATION = PHLTA,B249,441;%%
P.~Langacker,
{\it Precision Tests Of The Standard Model},
UPR-0435T
%\href{http://www.slac.stanford.edu/spires/find/hep/www?r=upr-0435t}{SPIRES entry}
{\it Invited talk given at PASCOS'90 Conf., Boston, MA, Mar 27-31, 1990};\br
%\AmaldiCN
U.~Amaldi, W.~de Boer and H.~F\"urstenau,
{\it Comparison of grand unified theories with 
electroweak and strong coupling constants measured at LEP},
Phys.\ Lett.\ B {\bf 260}, 447 (1991).
%%CITATION = PHLTA,B260,447;%%
}

%\GeorgiYF
\lref\GeorgiYF{
H.~Georgi and S.L.~Glashow,
{\it Unity Of All Elementary Particle Forces}, 
Phys.\ Rev.\ Lett.\  {\bf 32}, 438 (1974);
%%CITATION = PRLTA,32,438;%%
H.~Georgi, H.R.~Quinn and S.~Weinberg,
{\it Hierarchy Of Interactions In Unified Gauge Theories},
Phys.\ Rev.\ Lett.\  {\bf 33}, 451 (1974).
%%CITATION = PRLTA,33,451;%%
}

%\DerendingerHQ
\lref\DerendingerHQ{
J.P.~Derendinger, S.~Ferrara, C.~Kounnas and F.~Zwirner,
{\it On loop corrections to string effective field theories: 
Field dependent gauge couplings and sigma model anomalies},
Nucl.\ Phys.\ B {\bf 372}, 145 (1992).
%%CITATION = NUPHA,B372,145;%%
}

%\KaplunovskyRP
\lref\KaplunovskyRP{
V.S.~Kaplunovsky,
{\it One Loop Threshold Effects In String Unification},
Nucl.\ Phys.\ B {\bf 307}, 145 (1988)
[Erratum-ibid.\ B {\bf 382}, 436 (1992)], hep-th/9205068.
%%CITATION = HEP-TH 9205068;%%
}

%\IbanezZV
\lref\IbanezZV{
L.E.~Ibanez, D.~L\"ust and G.G.~Ross,
{\it Gauge coupling running in 
minimal SU(3) x SU(2) x U(1) superstring unification},
Phys.\ Lett.\ B {\bf 272} (1991) 251, hep-th/9109053;\br
%%CITATION = HEP-TH 9109053;%%
L.E.~Ibanez and D.~L\"ust,
{\it Duality anomaly cancellation, minimal string unification and the effective 
low-energy 
Lagrangian of 4-D strings},
Nucl.\ Phys.\ B {\bf 382}, 305 (1992), hep-th/9202046;\br
%%CITATION = HEP-TH 9202046;%%
P.~Mayr, H.P.~Nilles and S.~Stieberger,
{\it String unification and threshold corrections}, 
Phys.\ Lett.\ B {\bf 317} (1993) 53, hep-th/9307171;\br
%%CITATION = HEP-TH 9307171;%%
H.P.~Nilles and S.~Stieberger,
{\it How to Reach the Correct $\sin^2\theta_W$ and $\alpha_{\rm S}$ in 
String Theory},
Phys.\ Lett.\ B {\bf 367} (1996) 126, hep-th/9510009;
%%CITATION = HEP-TH 9510009;%%
{\it String unification, universal one-loop corrections and strongly coupled  heterotic 
string theory},
Nucl.\ Phys.\ B {\bf 499} (1997) 3, hep-th/9702110.
%%CITATION = HEP-TH 9702110;%%
}

%\AntoniadisQT
\lref\AntoniadisQT{
I.~Antoniadis, J.R.~Ellis, S.~Kelley and D.V.~Nanopoulos,
{\it The Price of deriving the Standard Model from string},
Phys.\ Lett.\ B {\bf 272}, 31 (1991).
%%CITATION = PHLTA,B272,31;%%
}

%\WittenMZ
\lref\WittenMZ{
E.~Witten,
{\it Strong Coupling Expansion Of Calabi-Yau Compactification},
Nucl.\ Phys.\ B {\bf 471}, 135 (1996), hep-th/9602070.
%%CITATION = HEP-TH 9602070;%%
}

%\DienesVG
\lref\DienesVG{
K.R.~Dienes, E.~Dudas and T.~Gherghetta,
{\it Grand unification at intermediate mass scales through extra dimensions},
Nucl.\ Phys.\ B {\bf 537} (1999)  47, hep-ph/9806292.
%%CITATION = HEP-PH 9806292;%%
}

%\AntoniadisEN
\lref\AntoniadisEN{
I.~Antoniadis, E.~Kiritsis and T.N.~Tomaras,
{\it A D-brane alternative to unification},
Phys.\ Lett.\ B {\bf 486} (2000) 186, hep-ph/0004214.
%%CITATION = HEP-PH 0004214;%%
}

\lref\rpradisi{G.~Pradisi, {\it Magnetized (Shift-)Orientifolds}, 
{\it The First International Conference on String Phenomenology},
World Scientific, hep-th/0210088.
%%CITATION = HEP-TH 0210088 ;%%
}

\lref\rkokora{C.~Kokorelis, {\it Deformed Intersecting D6-Brane GUTS I},
hep-th/0209202.
%%CITATION = HEP-TH  0209202;%%
}

\lref\rangles{M.~Berkooz, M.R.~Douglas and R.G.~Leigh, {\it Branes Intersecting
at Angles}, Nucl. Phys. B {\bf 480} (1996) 265, hep-th/9606139.
%%CITATION = HEP-TH 9606139;%%
}

\lref\rbgklnon{R.~Blumenhagen, L.~G\"orlich, B.~K\"ors and D.~L\"ust,
{\it Noncommutative Compactifications of Type I Strings on Tori with Magnetic
Background Flux}, JHEP {\bf 0010} (2000) 006, hep-th/0007024;
%%CITATION = HEP-TH 0007024;%%
{\it Magnetic Flux in Toroidal Type I Compactification}, Fortsch. Phys. 49
(2001) 591, hep-th/0010198.
%%CITATION = HEP-TH 0010198;%%
}

\lref\rangela{R.~Blumenhagen and C.~Angelantonj,
{\it Discrete Deformations in Type I Vacua},
Phys.Lett. {\bf B473} (2000) 86, hep-th/9911190.
%%CITATION = HEP-TH 9911190;%%
}

\lref\rdata{K. Hagiwara et al., {\it Review of Particle
Physics. Particle Data Group}, Phys. Rev. {\bf D66} (2002)  010001.}

\lref\rbgotwo{work in progress}

%\IbanezDJ
\lref\IbanezDJ{
L.E.~Ibanez,
{\it Standard Model Engineering with Intersecting Branes},
hep-ph/0109082.
%%CITATION = HEP-PH 0109082;%%
}

\lref\raads{C.~Angelantonj, I.~Antoniadis, E.~Dudas, A.~Sagnotti, {\it Type I
Strings on Magnetized Orbifolds and Brane Transmutation},
Phys. Lett. B {\bf 489} (2000) 223, hep-th/0007090;
%%CITATION = HEP-TH 0007090;%%
C.~Angelantonj, A.~Sagnotti, {\it Type I
Vacua and Brane Transmutation}, hep-th/0010279.
%%CITATION = HEP-TH 0001279;%%
}

\lref\rbkl{R.~Blumenhagen, B.~K\"ors and D.~L\"ust,
{\it Type I Strings with $F$ and $B$-Flux}, JHEP {\bf 0102} (2001) 030,
hep-th/0012156.
%%CITATION = HEP-TH 0012156;%%
}

\lref\rbbkl{R.~Blumenhagen, V.~Braun, B.~K\"ors and D.~L\"ust,
{\it Orientifolds of K3 and Calabi-Yau Manifolds with Intersecting D-branes},
JHEP {\bf 0207} (2002) 026, hep-th/0206038.
%%CITATION = HEP-TH 0206038;%%
}

\lref\rbbklb{R.~Blumenhagen, V.~Braun, B.~K\"ors and D.~L\"ust,
{\it The Standard Model on the Quintic}, hep-th/0210083.
%%CITATION = HEP-TH 0210083;%%
}

\lref\rura{A.M.~Uranga,
{\it Local models for intersecting brane worlds}, 
JHEP {\bf 0212} (2002) 058, hep-th/0208014.
%%CITATION = HEP-TH 0208014;%%
}

%\DixonPC
\lref\DixonPC{
L.J.~Dixon, V.~Kaplunovsky and J.~Louis,
{\it Moduli Dependence Of 
String Loop Corrections To Gauge Coupling Constants},
Nucl.\ Phys.\ B {\bf 355}, 649 (1991).
%%CITATION = NUPHA,B355,649;%%
}

\lref\PSph{G.K.~Leontaris and J.~Rizos,
{\it A Pati-Salam model from branes,}
Phys.\ Lett.\ B {\bf 510}, 295 (2001), hep-ph/0012255;\br
%%CITATION = HEP-PH 0012255;%%
L.L.~Everett, G.L.~Kane, S.F.~King, S.~Rigolin and L.T.~Wang,
{\it Supersymmetric Pati-Salam models from intersecting D-branes,}
Phys.\ Lett.\ B {\bf 531}, 263 (2002), hep-ph/0202100.
%%CITATION = HEP-PH 0202100;%%
}

\lref\KrauseGP{
A.~Krause,
{\it A small cosmological constant, grand unification and warped geometry},
hep-th/0006226.
%%CITATION = HEP-TH 0006226;%%
}

\lref\rcvetica{M.~Cvetic, G.~Shiu and  A.M.~Uranga,  {\it Three-Family
Supersymmetric Standard-like Models from Intersecting Brane Worlds}
Phys. Rev. Lett. {\bf 87} (2001) 201801,  hep-th/0107143.
%%CITATION = HEP-TH 0107143;%%
}

\lref\rcveticb{M.~Cvetic, G.~Shiu and  A.M.~Uranga,  {\it
Chiral Four-Dimensional N=1 Supersymmetric Type IIA Orientifolds from
Intersecting D6-Branes}, Nucl. Phys. B {\bf 615} (2001) 3, hep-th/0107166.
%%CITATION = HEP-TH 0107166;%%
}

\lref\rott{R.~Blumenhagen, B.~K\"ors, D.~L\"ust and T.~Ott, {\it
The Standard Model from Stable Intersecting Brane World Orbifolds},
Nucl. Phys. B {\bf 616} (2001) 3, hep-th/0107138.
%%CITATION = HEP-TH 0107138;%%
}

\lref\rbachas{C.~Bachas, {\it A Way to Break Supersymmetry}, hep-th/9503030.
%%CITATION = HEP-TH 9503030;%%
}

\lref\rafiruph{G.~Aldazabal, S.~Franco, L.E.~Ibanez, R.~Rabadan, A.M.~Uranga,
{\it Intersecting Brane Worlds}, JHEP {\bf 0102} (2001) 047, hep-ph/0011132.
%%CITATION = HEP-PH 0011132;%%
}

\lref\rafiru{G.~Aldazabal, S.~Franco, L.E.~Ibanez, R.~Rabadan, A.M.~Uranga,
{\it $D=4$ Chiral String Compactifications from Intersecting Branes},
J.\ Math.\ Phys.\  {\bf 42} (2001) 3103, hep-th/0011073.
%%CITATION = HEP-TH 0011073;%%
}

\lref\rimr{L.E.~Ibanez, F.~Marchesano, R.~Rabadan, {\it Getting just the
Standard Model at Intersecting Branes},
JHEP {\bf 0111} (2001) 002, hep-th/0105155.
%%CITATION = HEP-TH 0105155;%%
}

\lref\rkokoa{C.~Kokorelis, {\it GUT Model Hierarchies from Intersecting Branes},
JHEP {\bf 0208} (2002) 018, hep-th/0203187.
%%CITATION = HEP-TH 0203187;%%
}

\lref\rcls{M.~Cvetic, P.~Langacker, and G.~Shiu, {\it
 Phenomenology of A Three-Family Standard-like String Model},
Phys.Rev. {\bf D66} (2002) 066004, hep-ph/0205252;
%%CITATION = HEP-TH 0205252;%%
{\it  A Three-Family Standard-like Orientifold Model: 
Yukawa Couplings and Hierarchy},
Nucl.Phys. {\bf B642} (2002) 139, hep-ph/0206115.
%%CITATION = HEP-TH 0206115;%%
}

\lref\rquev{C.P.~Burgess, L.E.~Ibanez and F.~Quevedo, {\it
Strings at the Intermediate Scale, or is the Fermi Scale Dual to the Planck Scale?},
Phys.Lett. {\bf B447} (1999) 257, hep-th/9810535.
%%CITATION = HEP-TH 9810535;%%
}

\lref\rglift{S.~Kachru and J.~McGreevy, {\it
M-theory on Manifolds of $G_2$ Holonomy and Type IIA Orientifolds},
JHEP {\bf 0106} (2001) 027, hep-th/0103223.
%%CITATION = HEP-TH 0103223;%%
}

%\BlumenhagenUA
\lref\BlumenhagenUA{
R.~Blumenhagen, B.~K\"ors, D.~L\"ust and T.~Ott,
{\it Hybrid Inflation in Intersecting Brane Worlds},
Nucl.Phys. {\bf B641} (2002) 235, hep-th/0202124.
%%CITATION = HEP-TH 0202124;%%
}

%\KlebanovMY
\lref\KlebanovMY{
I.R.~Klebanov and E.~Witten,
{\it Proton decay in intersecting D-brane models}, hep-th/0304079.
%%CITATION = HEP-TH 0304079;%%
}

%\BlumenhagenGW
\lref\BlumenhagenGW{
R.~Blumenhagen, L.~G\"orlich and T.~Ott,
{\it Supersymmetric intersecting branes on the type IIA 
$T^6/Z(4)$  orientifold}, JHEP {\bf 0301} (2003) 021,
hep-th/0211059.
%%CITATION = HEP-TH 0211059;%%
}

%\CremadesVA
\lref\CremadesVA{
D.~Cremades, L.E.~Ibanez and F.~Marchesano,
{\it Towards a theory of quark masses, mixings and CP-violation},
hep-ph/0212064;
%%CITATION = HEP-PH 0212064;%%
{\it Yukawa couplings in intersecting D-brane models},
hep-th/0302105.
%%CITATION = HEP-TH 0302105;%%
}

%\LustKY
\lref\LustKY{
D.~L\"ust and S.~Stieberger,
{\it Gauge threshold corrections in intersecting brane world models},
hep-th/0302221.
%%CITATION = HEP-TH 0302221;%%
}

\lref\TW{
T.~Friedmann and E.~Witten,
{\it Unification scale, proton decay, and manifolds of G(2) holonomy,} hep-th/0211269.
%%CITATION = HEP-TH 0211269;%%
}

%\AbelVV
\lref\AbelVV{
S.A.~Abel and A.W.~Owen,
{\it Interactions in intersecting brane models},
hep-th/0303124;
%%CITATION = HEP-TH 0303124;%%
S.A.~Abel, M.~Masip and J.~Santiago,
{\it Flavour changing neutral currents in intersecting brane models},
hep-ph/0303087.
%%CITATION = HEP-PH 0303087;%%
}

%\AbelFK
\lref\AbelFK{
S.A.~Abel, M.~Masip and J.~Santiago,
{\it Flavour changing neutral currents in intersecting brane models},
hep-ph/0303087.
%%CITATION = HEP-PH 0303087;%%
}

%\HoneckerVQ
\lref\HoneckerVQ{
G.~Honecker,
{\it Chiral supersymmetric models on an orientifold of 
Z(4) x Z(2) with  intersecting D6-branes},
hep-th/0303015.
%%CITATION = HEP-TH 0303015;%%
}

%\DienesDU
\lref\DienesDU{
K.R.~Dienes,
{\it String Theory and the Path to Unification: 
A Review of Recent Developments},
Phys.\ Rept.\ {\bf 287} (1997) 447 , hep-th/9602045.
%%CITATION = HEP-TH 9602045;%%
}

%\CveticXS
\lref\CveticXS{
M.~Cvetic and I.~Papadimitriou,
{\it More supersymmetric standard-like models from intersecting 
D6-branes on  type IIA orientifolds},
hep-th/0303197.
%%CITATION = HEP-TH 0303197;%%
}

%\CveticCH
\lref\CveticCH{
M.~Cvetic and I.~Papadimitriou,
{\it Conformal field theory couplings for intersecting 
D-branes on  orientifolds},
hep-th/0303083.
%%CITATION = HEP-TH 0303083;%%
}

%%% Title page
\Title{\vbox{
 \hbox{HU--EP-03/22}
 \hbox{hep-th/0305146}}}
 %\vskip-1cm
{\vbox{\centerline{Gauge Unification in Supersymmetric}
\vskip 0.3cm \centerline{Intersecting Brane Worlds}
}}
\centerline{Ralph Blumenhagen{}, Dieter L\"ust{} and Stephan Stieberger }
\bigskip\medskip
\centerline{ {\it Humboldt-Universit\"at zu Berlin, Institut f\"ur
Physik,}}
\centerline{\it Newtonstra{\ss}e 15, 12489 Berlin, Germany}
\centerline{\tt Email:
blumenha, luest, stieberg@physik.hu-berlin.de}
\bigskip
\bigskip

\centerline{\bf Abstract}
\noindent

We show that contrary to first expectations  realistic three generation
supersymmetric intersecting brane world models give rise to phenomenologically 
interesting predictions
about gauge coupling unification. Assuming  the most economical
way of realizing the matter content
of the MSSM via intersecting branes
we obtain a model independent
relation among the three gauge coupling constants at
the string scale.
In order to correctly reproduce the experimentally known 
values of $\sin^2\theta_w(M_z)$ and $\alpha_s(M_z)$
this relation leads to natural gauge coupling
unification at a string scale 
close to the standard GUT scale $2\times 10^{16}~{\rm GeV}$.
Additional vector-like matter can push the  unification scale
up to the Planck scale.

%\medskip

\Date{05/2003}
%%% text
\newsec{Introduction}

The experimental observation \EllisZQ\ that all three momentum
dependent gauge coupling constants
of the minimal supersymmetric Standard Model (MSSM),
although being rather different at the 
low energy scale $M_Z$, apparently meet at a 
high mass scale $M_X$ ($M_X\gg M_Z$) is very striking and
naturally calls for a convincing theoretical explanation,
commonly called
{\sl Grand Unification}. This means that above $M_X$ all interactions of
the Standard Model should be unified into a single
theory, which predicts the observed relations among the coupling constants
in a natural manner. 
At the one-loop level, the gauge couplings of the low energy
Standard Model gauge theory
evolve as
\eqn\rgg{{4\pi\over g_a^2(\mu)}=k_a{4\pi\over g^2_X
}+{b_a\over 2\pi} \log\left({\mu\over M_X}\right)+\Delta_a }
with $a=\lbrace SU(3),SU(2),U(1)_Y\rbrace$.
Here, the $b_a$ are the renormalization group coefficients which can be
computed by knowing the charges of the particle spectrum 
with respect to the three gauge groups.
The $k_a$ are constants which parameterize the tree level relations
among the gauge couplings at the unification scale $M_X$, and which should
be predicted completely or at least in
part by the GUT theory. 
Finally, the quantities $\Delta_a$ are the 
one-loop group dependent threshold corrections due
to heavy states at the scale $M_X$. They should also be calculable in any given GUT theory.
Therefore without further input these
three equations contain as parameters the three experimentally known
couplings $g_a(M_Z)$ and the a priori
unknown parameters $M_X$ and $g_X$.

Starting
from the measured low-energy data, namely from $\alpha_3(M_Z)$,
$\alpha(M_Z)$ and  $\sin^2\theta_w(M_Z)$,
and extrapolating to high energies by the use of the 
spectrum of the MSSM in the renormalization group
equations \rgg, it turns out that at a scale $M_X\simeq 
2\cdot 10^{16}~{\rm GeV}$ the constants $k_a$ obey $k_3\simeq k_2\simeq
{3\over 5} k_1$
to a surprising level of accuracy. Therefore this scale suggests itself as 
a natural unification scale. However it us useful to emphasize 
that different unification scales are needed if the GUT theory provides
different relations among the $k_a$.

In supersymmetric field theory, GUT theories \GeorgiYF\
with unifying gauge group
$G\supset SU(3)\times SU(2)\times U(1)_Y$ provide a beautiful
theoretical framework.
In the case of $G=SU(5)$ there are indeed
two fixed relations among the gauge
couplings at $M_X$ provided by the group structure of $SU(5)$, namely
$k_1={5\over 3}$, $k_2=k_3=1$.
This implies that with $\alpha_3(M_Z)\simeq 0.12$ and
$\alpha(M_Z)\simeq 1/128$, and assuming that the one-loop threshold corrections
are small in field theory, 
two renormalization group equations fix $M_X\simeq 
2\cdot 10^{16}~{\rm GeV}$ and $\alpha_X\simeq 1/24$; 
then the remaining equation provides a prediction
for the coupling $\sin^2\theta_w(M_Z)$.
Using the spectrum of the MSSM, the
$SU(5)$ GUT theory so far is in excellent agreement with the experimental
measurements. This
is depicted in figure 1 (the upper curve denotes the running of
${3\over 5}\alpha_Y^{-1}$).
\fig{GUT unification}{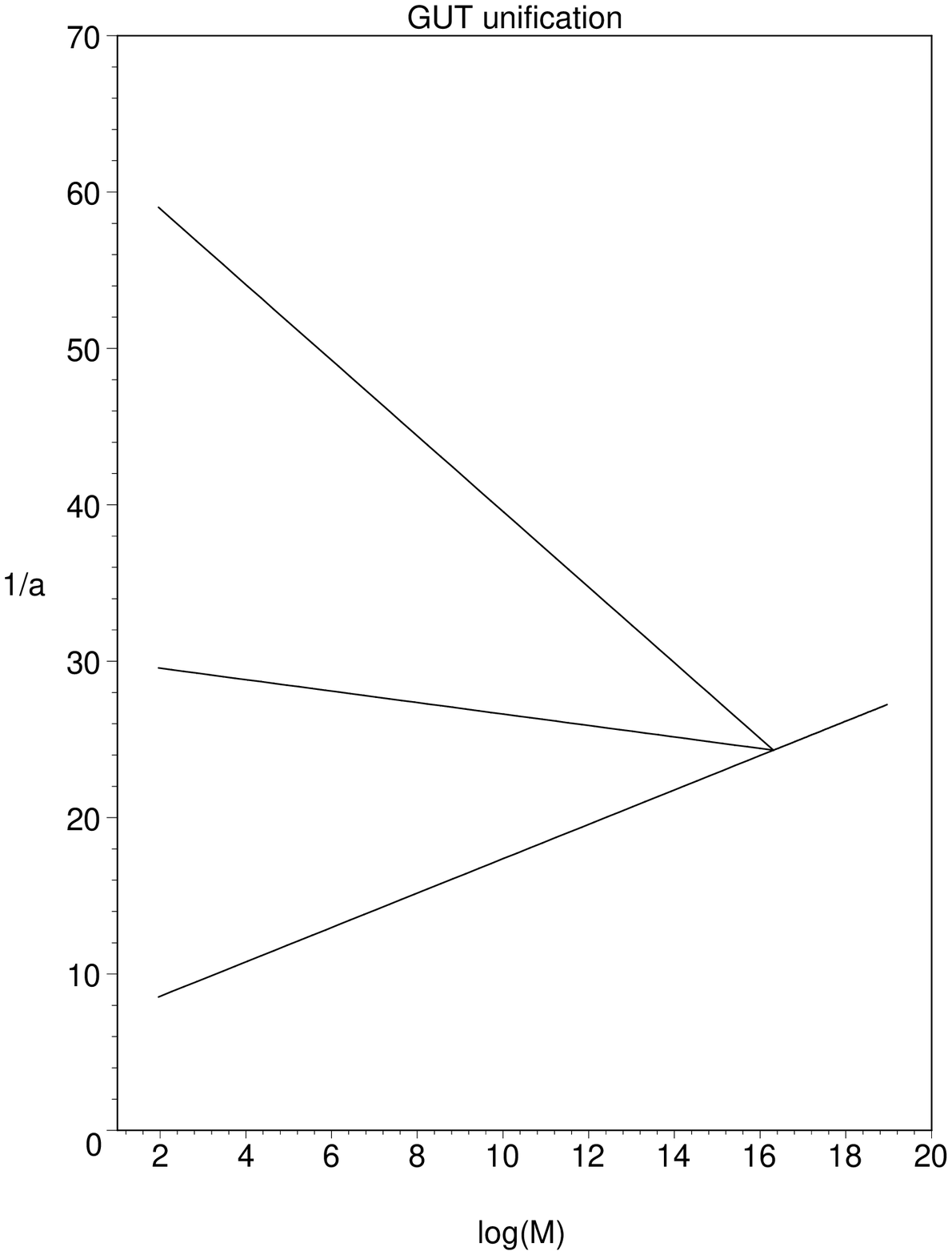}{6truecm}

%\noindent
String theoretical {\sl Grand Unification} is more ambitious than field theory,
since it attempts to unify gravity with the gauge interactions.
Therefore one expects that in string theory unification takes place at
the string scale $M_s$, i.e. $M_X=M_s$. This scale
should be related to
the Planck mass $M_{pl}=G_N^{-1/2}
\simeq 1.2\cdot 10^{19}~{\rm GeV}$ in one way or the other.
Moreover
at $M_s$  all gauge couplings should be expressible in terms of the
string coupling constant $g_{st}$, making the constants $k_a$ in principle
calculable.
Perturbative heterotic string compactifications \foot{See \DienesDU\
for a review.} 
provide a very concrete
realization of string gauge coupling unification.
Here
the constants $k_a$ are just the Kac-Moody levels of the corresponding
gauge Kac-Moody algebras \GinspargEE, where the most favorite  ($SU(5)$ like)
choice is $k_3=k_2={3\over 5}k_1=1$ (however see also the discussion
in \IbanezBD).
Furthermore, the heterotic string scale $M_s$ can be related to the Planck mass
as follows \doubref\KaplunovskyRP\DerendingerHQ
\eqn\num{
M_s\simeq g_{st}\cdot 0.058\cdot  M_{pl}
\, .}
This relation eliminates one parameter from 
the three eqs.\rgg, so knowing $\alpha(M_Z)$
one can predict both $\alpha_3(M_Z)$ and $\sin^2\theta_w(M_Z)$.
Assuming the spectrum of the MSSM and neglecting one-loop threshold corrections
the heterotic string prediction is off the experimental values
for these two couplings by several standard deviations. 
In other words, with $g_{st}\simeq 0.7$ the heterotic string scale becomes
too large, namely
$M_s\simeq 5\cdot 10^{17}~{\rm GeV}$.
As being roughly of the same size as the 
unification  scale of $2\times 10^{16}
~{\rm GeV}$, where the three coupling constants of the MSSM
meet according to their 
experimental values, this was considered to be an encouraging result
for heterotic string gauge unification. Nevertheless,
the remaining discrepancy between the unification scale
and $M_s$ is too large in order
to be neglected. Therefore the influence of extra states
at intermediate mass scales \AntoniadisQT\ was discussed 
in order to improve the
situation.

Alternatively the effect of one-loop
string threshold corrections $\Delta_a$
\threeref\KaplunovskyRP\DixonPC\IbanezZV\
due to heavy string modes was investigated.
The quantities $\Delta_a$ depend in general on the gauge group
and induce a deviation of the relations among the gauge couplings at $M_s$
from their tree level values
(absorbing this effect into the $k_a$, one gets sort of effective one-loop
Kac-Moody levels $k_a^{1-loop}$).
In addition, in \WittenMZ\
it was shown that in heterotic M-theory compactifications
the string scale could be lowered such that agreement with the experiment
is in principle possible.
However the predictive power of the heterotic string gauge
unification is somehow weakened  by all these considerations.

More recently, gauge coupling unification was also discussed in the framework
of non-supersymmetric large extra dimension scenarios with a string
scale in the TeV region.
In one proposal one uses polynomial running of the gauge couplings \DienesVG,
whereas the second approach outlined in \AntoniadisEN\  uses ideas similar
in spirit to what we are to present in the following (see also \KrauseGP).

In this paper we want to investigate the question in how far 
intersecting brane world (IBW) models
\refs{\rbgklnon\raads\rafiru\rafiruph\rbkl\rimr
\rott\rcvetica\rcveticb\rkokoa\rcls
\rbbkl\rura\rpradisi\BlumenhagenGW\LustKY\HoneckerVQ\CveticCH\AbelVV\CveticXS
-\KlebanovMY}
can predict the values of the low-energy gauge couplings in a
satisfactory way.
The main new ingredient in these models is that they contain
intersecting D-branes and open strings in a consistent manner
providing  simple mechanisms to generate non-Abelian gauge interactions
and
chiral fermions
in a systematic way. To be specific we will consider type IIA 
orientifolds with several stacks of $D6_a$-branes, each being wrapped
around individual compact homology 3-cycles $\pi_a$ of the internal space.
Hence the effective open string gauge theories with
groups $G_a$ live in the 7-dimensional subspaces ${\IR}^{1,3}\otimes
\pi_a$. The chiral matter fields are located at the intersection points
of the D6-branes, which are points on the internal space, i.e.
the chiral matter fields live just in ${\IR}^{1,3}$.
In this way one can systematically build intersecting brane world models
which come remarkably close to the (supersymmetric) Standard Model.

In contrast to the heterotic string, here
each gauge factor comes with its own gauge coupling,
which at string tree-level can be deduced from the
Dirac-Born Infeld action. Here we simply state
the well known result that the tree level gauge coupling 
at the string scale is given
by\foot{A very explicite derivation is for instance contained in \KlebanovMY.}
\eqn\gauge{   {4\pi \over g_a^2}={M_s^3\, V_{a} \over (2\pi)^3\, g_{st}\, 
\kappa_a,}}
where $V_{a}$ denotes the volume of the 3-cycle the $D6$-branes are wrapped
on. The extra factor $\kappa_a$ is related to the  difference between  the
gauge couplings for $U(N_a)$ and $SP(2N_a)/SO(2N_a)$ branes, 
namely $\kappa_a=1$ for $U(N_a)$ and $\kappa_a=2$ for $SP(2N_a)/SO(2N_a)$.
Note, that in deriving \gauge\ one has normalized the gauge fields in the
canonical way, i.e. $A_\mu=A^a_\mu\, T_a$ with ${\rm tr}(T_a\, T_b)=1/2$.
We see that by setting $g_{st}=g_X$ the constants $k_a$ in eq.\rgg\
can by identified as 
\eqn\kas{
k_a={M_s^3\, V_a\over (2\pi)^3\, \kappa_a}\, .}
The constants $k_a$
now depend on the internal volumes $V_a$
and since they are in general different, the $k_a$ are 
generically independent and different.
Hence gauge coupling unification does not seem to occur in type II
brane world models in a very natural way.

By dimensionally reducing the type IIA gravitational action one can  
similarly express the Planck mass in terms of stringy parameters
\eqn\grav{   {M_{pl}^2}={8\, M_s^8\, V_{6} \over (2\pi)^6\, g^2_{st}},}
where $V_6$ is the overall volume of the Calabi-Yau manifold.
The two equations \gauge\ and \grav\ can be used to eliminate
the unknown string coupling constant
\eqn\elim{   {1\over \alpha_a}={M_{pl} \over 2\sqrt{2}\, \kappa_a\, M_s} 
                      {V_a\over \sqrt{V_6}}  .}
Let us point out, that in that form the gauge coupling $\alpha_a$ depends on 
the complex structure moduli only. On the other hand the size of the 
Planck mass \grav\ is governed by the overall volume $V_6$.

Since, as already emphasized, each gauge coupling depends
on the size of the  3-cycle the $D6$ is wrapped on, one might think
that all gauge couplings are in general independent and 
gauge unification does not give very restrictive constraints. 
In the following section we will show that in a model
independent bottom up approach the three phenomenological
requirements, namely
of (i) that the standard model branes mutually preserve
${\cal N}=1$ supersymmetry, of (ii)
realizing a 3-generation  MSSM like model from intersecting branes,
i.e. realizing the intersection numbers of a 3 generation MSSM, and
of (iii) getting a massless $U(1)_Y$ gauge boson with the correct 
hypercharges of the matter fields,
lead in a natural way to one non-trivial relation 
among the tree level gauge couplings.
Against our first intuition this relation
allows for natural gauge coupling unification and makes
a prediction about the string scale in 
this class of supersymmetric intersecting brane world models.

\newsec{Realizing MSSM like models in intersecting brane worlds}

The idea of intersecting brane worlds can be described very simple. 
One starts with an orientifold of Type IIA string theory on
a Calabi-Yau space. The anti-holomorphic orientifold projection
introduces $O6$ planes in the background, which are wrapped on some
of the 3-cycles $\pi_{O6}$ 
 of the Calabi-Yau. These orientifold planes are charged under
the Type IIA R-R 7-form and due to its tension also couple to the 
gravitational field.  
Since on a compact background the sum of all charges must vanish,
one has to introduce $D6$-branes into  the background as well, so
that the overall 7-form charge vanishes. In general it is 
not necessary to place these $D6$-branes right on top
of the orientifold planes, instead they can in general wrap
different 3-cycles $\pi_a$  of the underlying Calabi-Yau manifold. 
The condition that the overall 7-form charge vanishes 
gives rise to the following R-R tadpole cancellation condition,
\eqn\tadhom{ \sum_a  N_a\, (\pi_a + \pi'_a)-4\, \pi_{O6}=0, }
where $\pi'_a$ denotes the orientifold image of the cycle
$\pi_a$. 

In general such an arrangement will break supersymmetry, but
non-trivial models have been explicitly constructed, where
an ${\cal N}=1$ supersymmetry is preserved by all $D6$ branes.
Moreover, it has been pointed out that at strong coupling 
such supersymmetric models do lift to M-theory compactifications on singular 
$G_2$-manifolds \refs{\rglift,\rcveticb}.  

 From the phenomenological point of view such string models 
are interesting, as they allow to search systematically 
for stringy realizations of the standard model respectively the 
MSSM. The gauge degrees of freedom are confined to the
world volume of the $D6$-branes.
In particular each stack of D6-branes wrapping a 3-cycle, which is
not invariant under the orientifold projection gives rise
to a gauge factor $U(N_a)$. If however the 3-cycle is invariant
one can also get gauge groups $SP(2N_a)$ and $SO(2N_a)$.
The chiral matter fields
are localized at the four-dimensional intersection locus  
between two $D6$-branes. The number of such intersections is 
given by the topological intersection number between the two
different 3-cycles the $D6$-branes are wrapped on.
Therefore, such models naturally give rise to chirality
and family replication. 
The general chiral spectrum in terms of the topological intersection numbers 
is  shown in Table 1 \rbbkl. 
\vskip 0.8cm
\vbox{ \centerline{\vbox{ \hbox{\vbox{\offinterlineskip
\def\tablespace{height2pt&\omit&&
 \omit&\cr}
\def\tablerule{\tablespace\noalign{\hrule}\tablespace}

\hrule\halign{&\vrule#&\strut\hskip0.2cm\hfill #\hfill\hskip0.2cm\cr
& Representation  && Multiplicity &\cr
\tablerule
& $[{\bf A_a}]_{L}$  && ${1\over 2}\left(\pi'_a\circ \pi_a+\pi_{{\rm O}6} \circ \pi_a\right)$   &\cr
\tablerule
& $[{\bf S_a}]_{L}$
     && ${1\over 2}\left(\pi'_a\circ \pi_a-\pi_{{\rm O}6} \circ
\pi_a\right)$   &\cr \tablerule & $[{\bf (\o N_a,N_b)}]_{L}$  &&
$\pi_a\circ \pi_{b}$ &\cr \tablerule & $[{\bf (N_a, N_b)}]_{L}$ &&
$\pi'_a\circ \pi_{b}$   &\cr 
}\hrule}}}} 
\centerline{ \hbox{{\bf
Table 1:}{\it ~~ Chiral spectrum}}} } 
\vskip 0.5cm
\noindent

So far concrete intersecting brane world models have been constructed
on the torus $T^6$, on toroidal orbifolds and also on the
quintic Calabi-Yau manifold \rbbklb. 
In particular,
supersymmetric standard-like models have been studied on the
$T^6/\ZZ_2\times \ZZ_2$ \refs{\rcveticb,\rpradisi,\CveticXS}, 
$T^6/\ZZ_4$ \BlumenhagenGW\ and  $T^6/\ZZ_4\times \ZZ_2$ \HoneckerVQ\
orbifold. However, we think it is fair to say that none
of these models are completely realistic yet. 
However, there are some general patterns supersymmetric standard-like
models apparently share. 
In the following our bottom up analysis will not rely on any concrete 
model, but will only make use of these general properties
which follow from the three phenomenological requirements spelled out at the
end of the introduction.

First there are so far two simple ways to embed the standard model
gauge group into products of unitary and symplectic gauge groups.
Both of them use four stacks of $D6$-branes, which give rise to the
initial gauge symmetries
\eqn\gaugein{\eqalign{ A&:\  U(3)\times SP(2) \times U(1)\times U(1) \cr
                       B&:\  U(3)\times U(2) \times U(1)\times U(1). \cr }} 
The difference is that in the first example \CremadesVA\ the $SU(2)_L$ symmetry
of the standard model is realized directly as an $SP(2)=SU(2)$ 
gauge group on the branes. In the second case one embeds $SU(2)_L$
into the $U(2)$ gauge factor \refs{\rbkl,\rimr}. Since the anomaly
conditions are less constraining for the class $A$ and the Standard
Model embedding is more natural, let us  explore  this
example in more detail. 

Second, the chiral spectrum of the intersecting brane world
model should be identical to the chiral spectrum 
of the standard model particles.
This fixes uniquely the intersection numbers of the 
3-cycles, $(\pi_a,\pi_b,\pi_c,\pi_d)$, 
the four stacks of $D6$-branes are wrapped on.
\vskip 0.8cm 
\vbox{ \centerline{\vbox{
\hbox{\vbox{\offinterlineskip
\def\tablespace{height2pt&\omit&&\omit&&\omit&&
 \omit&\cr}
\def\tablerule{\tablespace\noalign{\hrule}\tablespace}

\hrule\halign{&\vrule#&\strut\hskip0.2cm \hfill #\hfill\hskip0.2cm\cr
& field && sector && I  && $SU(3)\times SU(2)\times U(1)_a\times U(1)_c
\times U(1)_d$   &\cr
\tablerule
& $q_L$ && (ab) && 3 && $(3,2;1,0,0)$ &\cr
& $u_R$  && (ac)  && 3 && $(\o{3},1;-1,1,0)$ &\cr
& $d_R$  && (ac') && 3 && $(\o{3},1;-1,-1,0)$ &\cr
\tablerule
& $e_L$  && (db) && 3 && $(1,2;0,0,1)$ &\cr
& $e_R$  && (dc') && 3 && $(1,1;0,-1,-1)$ &\cr
& $\nu_R$  && (dc)&& 3 && $(1,1;0,1,-1)$ &\cr
}\hrule}}}}
\centerline{
\hbox{{\bf Table 2:}{\it ~~ Chiral spectrum for the A model}}}
}
\vskip 0.5cm
Note, that the $U(1)\subset U(N)$ gauge factors in Table 2 are not canonically 
normalized. The canonically normalized ones are given by
$\widetilde{U(1)}=U(1)/\sqrt{2\, N}$. 
The hypercharge $Q_Y$ is given as the following linear combination
of the three $U(1)$s
\eqn\hyper{   Q_Y={1\over 3}Q_a-Q_c-Q_d .}
As first described in \rimr\ in general some of the stringy $U(1)$s are
anomalous and get a mass via some generalized Green-Schwarz mechanism.
However, for intersecting brane worlds it can also happen that via
axionic couplings  some anomaly-free abelian gauge groups become massive. 

The condition that a linear combination $U(1)_Y=\sum_i c_i U(1)_i$
remains massless reads \rbbkl
\eqn\green{    \sum_i  c_i\, N_i\, \left(\pi_i-\pi_i'\right)=0 .}
In general, if the hypercharge is such a linear combination of $U(1)$s,
$Q_Y=\sum_i c_i Q_i$, then the gauge coupling is given by
\eqn\hgauge{   {1\over \alpha_Y}=\sum_i {N_i\, c_i^2\over 2}
                           {1\over \alpha_i}, }
where we have taken into account that the $U(1)$s in Table 2 
are not canonically normalized.
Therefore, in our case the gauge coupling of the hypercharge
is given as
\eqn\hgaugebb{   {1\over \alpha_Y}={1\over 6}{1\over \alpha_a} +
                                  {1\over 2}{1\over \alpha_c} +
                                  {1\over 2}{1\over \alpha_d} .}
Since naively one would guess that $\alpha_c$ and $\alpha_d$ 
are independent parameters one cannot derive any definite low-energy
prediction from this formula. However, there exists
a most natural and economical way of realizing the Standard Model intersection
numbers.

Say one finds two supersymmetric
3-cycles $\pi_a$ and $\pi_b$ with the intersection
numbers $\pi_a\circ \pi_b=3$ then homologically choosing $\pi_d=\pi_a$
gives the right intersection numbers for $\pi_d$.
Therefore the two volumes $V_a$ and $V_d$ have to agree:
$V_a=V_d$.
This also follows from the fact that the gauge symmetry can be enhanced
$U(3)_a\times U(1)_d\rightarrow U(4)$, if we put the brane d on top of the
branes a.
The condition \green\ that $U(1)_Y$ remains massless simply implies homologically 
$\pi_c'=\pi_c$, i.e. by really placing $\pi_c$ on top of an orientifold
plane one can enhance the gauge group for $\pi_c$ from $U(1)$ to
$SP(2)$. Therefore, at the bottom of this simple realization
there lies an extended Pati-Salam like model\foot{Gauge unification 
in  Pati-Salam like D-brane models has been discussed in Refs. \PSph.} 
discussed $U(4)\times SU(2)\times SU(2)$.
On the one hand, from the field theory point of view
it is very natural to assume, 
that the two gauge couplings of the
two $SU(2)$ factors have the same gauge coupling, i.e. there is 
an additional $\ZZ_2$ symmetry.\foot{This  $\ZZ_2$ symmetry
could be understood in some decompactification limit where
$SP(2)_b\times SP(2)_c$ might be unified to $SP(4)$.}
On the other hand,  from the stringy point of view, even though we cannot
rigorously prove it in the general case, the constraints from supersymmetry,
i.e. the requirement that all branes are calibrated in the same way,
and the intersection numbers $\pi_a\circ\pi_b=-\pi_a\circ\pi_c=3$
do not  seem to leave very much room to evade that the internal
volumes for the cycles $\pi_b$ and $\pi_c$  agree: $V_b=V_c$.
Following these arguments it follows that
$\alpha_d=\alpha_a=\alpha_s$ and
$\alpha_c={1\over 2}\alpha_b={1\over 2}\alpha_w$. 
In all known concrete supersymmetric models, 
like the one discussed in   \CremadesVA,
these relations among the gauge couplings hold
\foot{In toroidal models with
four stacks of 
$D6_a$ branes ($a=1,\cdots ,4$) 
wrapped on non--trivial 
$3$--cycles of a six--dimensional torus $T^6=\prod\limits_{j=1}^3 T^{2,j}$
with the three complex structures $U^j=\fc{R_2^j}{R_1^j}$ and wrapping numbers
$(n^j_a,m^j_a)$
one obtains for the gauge couplings for each stack $a$ 
\eqn\respgauge{\eqalign{
\alpha^{-1}_a&=\fc{1}{2\sqrt{2}\kappa_a}\ \fc{M_{pl}}{M_s}\cr 
&\times\lf(n_a^1n_a^2n_a^3 \fc{1}{\sqrt{U^1U^2U^3}}-
n_a^1m_a^2m_a^3\sqrt{\fc{U^2U^3}{U^1}}-m_a^1n_a^2m_a^3\sqrt{\fc{U^1U^3}{U^2}}-
m_a^1m_a^2n_a^3\sqrt{\fc{U^1U^2}{U^3}}\ri) }}
in the case of a supersymmetric cycle, {\it i.e.} 
$\sum\limits_{j=1}^3 \arctan\lf( \fc{m_a^j R_2^j}{n^j_a R_1^j} \ri)=0$.
A possible choice for the wrapping numbers 
that fulfills the conditions worked out before
is
 \CremadesVA:
$\pi_a=\{(1,0),(1/\rho,3\rho),(1/\rho,-3\rho)\}$, 
$\pi_b=\{(0,1),(1,0),(0,-1)\}$, 
$\pi_c=\{(0,1),(0,-1),(1,0)\}$, 
$\pi_d=\pi_a$, with $N_a=3$,\ $N_b=N_c=N_d=1$.
${\cal N}=1$ supersymmetry requires 
$U^2=U^3$ and allows for the two possibilities $\rho=1,1/3$.
    From \respgauge\ and \hyper\ 
we obtain for the three gauge couplings ($\kappa_b=2$)
at the string scale:
\eqn\system{\eqalign{
\alpha^{-1}_s&=\fc{1}{2\sqrt{2}}\ \fc{M_{pl}}{M_s}\ \lf(
\fc{1}{\rho^2}\fc{1}{\sqrt{U^1} U^2}+9\rho^2\fc{U^2}{\sqrt{U^1}}\ri)\ ,\cr
\alpha^{-1}_w&=\fc{1}{2\sqrt{2}}\ \fc{M_{pl}}{M_s}\ \fc{\sqrt{U^1}}{2}\ ,\cr
\alpha^{-1}_Y&=\fc{1}{2\sqrt{2}}\ \fc{M_{pl}}{M_s}\ \lf[\fc{2}{3}\lf(
\fc{1}{\rho^2}\fc{1}{\sqrt{U^1} U^2}+
9\rho^2\fc{U^2}{\sqrt{U^1}}\ri) +\h \sqrt{U^1}\ri]\ .}}
Immediatley, we see, that the condition (2.9) is satisfied.}.

This natural choice leads to a non-trivial relation
among the three Standard Model gauge couplings at the string scale
\eqn\hgaugeb{   {1\over \alpha_Y}={2\over 3}{1\over \alpha_{s}} +
                                  {1\over \alpha_{w}}  .}
Note, that this relation is compatible with the $SU(5)$ GUT prediction
\eqn\gut{  {1\over \alpha_{s}}={1\over \alpha_{w}}={3\over 5}{1\over
\alpha_Y,}}
even though it is less constraining. It is amusing that even though there
seems to be no hidden $SU(5)$ in our construction, we derived 
a compatible relation from intersecting branes.
Note, that the relation for the internal volumes of the four D6-branes,
$V_a=V_d$ and $V_b=V_c$, and in consequence the relation
\hgaugeb\ 
might also hold for IBW models, even when there is no point in moduli
space where the gauge symmetry is enhanced to $U(4)\times SU(2)\times
SU(2)$. In fact, all the results we are going to derive in the following
are true as long as \hgaugeb\ holds.   

By making similar natural assumptions for the Standard Model
realization of type $B$ in \gaugein, one can derive the same
relation \hgaugeb, where again at the bottom there lies an
enhanced $U(4)\times U(2)\times U(2)$ Pati-Salam-like IBW model.

\newsec{Running of the gauge couplings}

Using the string prediction of the relation among  the gauge couplings 
at the string scale, we can now use the one-loop running of the
gauge couplings down to the weak scale. As a first step we are 
ignoring string threshold corrections, which for concrete models
may be determined from the results derived in \doubref\TW\LustKY.

In the absence
of threshold corrections,
the one-loop running of the three gauge couplings is described
by the well known formulas
\eqn\running{\eqalign{ {1\over \alpha_s(\mu)}&={1\over \alpha_s} + {b_3\over 2\pi} 
                   \ln\left({\mu\over M_s}\right) \cr
                   {\sin^2\theta_w(\mu)\over \alpha(\mu)}&={1\over \alpha_w} + 
                        {b_2\over 2\pi} 
                   \ln\left({\mu\over M_s}\right) \cr
                   {\cos^2\theta_w(\mu)\over \alpha(\mu)}&={1\over \alpha_Y} + 
                        {b_1\over 2\pi} 
                   \ln\left({\mu\over M_s}\right) \cr}}
where $(b_3,b_2,b_1)$ are the one-loop beta-function coefficients for $SU(3)_c$,
$SU(2)_L$ and $U(1)_Y$.
Remember that for the matter spectrum of the MSSM these coefficients
are given by $(b_3,b_2,b_1)=(3,-1,-11)$ \foot{Note, that the concrete model
in \CremadesVA\
has additional non-chiral matter, in particular in the adjoint
representations spoiling asymptotic freedom of $SU(3)_c$.}.
Using the relation \hgaugeb\ at the string scale yields 
\eqn\rela{ {2\over 3}{1\over \alpha_{s}(\mu)}+{2\sin^2\theta_w(\mu)-
                1 \over \alpha(\mu)}={B\over 2\pi}\, 
     \ln\left({\mu\over M_s}\right) .}
with
\eqn\bbb{   B={2\over 3}\,b_3+ b_2-b_1.}
Therefore, once a concrete string model is given, one can compute 
beta-function coefficients and use \rela\ to compute the unification scale by
inserting the measured values of the couplings constants at the weak scale.
In the following we use the following values for the Standard Model parameters
taken from \rdata
\eqn\values{\eqalign{  M_Z&=91.1876(21)\ {\rm GeV},\quad \alpha_s(M_Z)=0.1172(20), \cr
                     \ \alpha(M_Z)&={1\over
127.934(27)}, \quad \, \, \sin^2\theta_w(M_Z)=0.23113(15) .\cr}}
It is clear from \rela\ that the resulting value of the unification scale
 only depends
on the combination $B$ of the beta-function coefficients. Therefore, there
exists a whole class of 3-generation supersymmetric intersecting brane 
world models
with additional non-chiral matter that lead to the same prediction for the
string scale. We will come back to this point in section 4. 

Now, assuming the matter spectrum of the MSSM
we get $B=12$ and the resulting  value for the unification
 scale turns out to be
\eqn\wow{   M_X=2.04\cdot 10^{16}\, {\rm GeV} .}
Since in \grav\ we still have the internal volume as an unfixed parameter,
in contrast to the heterotic string, we can identify the
string scale with the unification scale $M_s=M_X$.

Of course, for
the individual gauge couplings at the string scale we get
\eqn\couplstr{  \alpha_s(M_s)=\alpha_w(M_s)={5\over 3}\alpha_Y(M_s)=0.041,}
which are just the supersymmetric GUT scale values with the Weinberg angle
being $\sin^2\theta_w(M_s)=3/8$.
We conclude that the described class
of realistic intersecting brane world models, under the assumption
of a matter content with the same $B$ coefficient as the MSSM matter content
features perturbative "gauge coupling
unification" at the GUT scale. Note, that the string prediction is weaker
than the GUT prediction, as the  latter leads to  two conditions among the 
gauge couplings at the GUT scale, which allows to derive one non-trivial
relation among the three couplings at the weak scale.
In figure 2 we have depicted the running of the three gauge couplings,
where for illustrative purposes we have shown the $SU(5)$ coupling
$\alpha'={5\over 3}\, \alpha_Y$.  

In order to match the values of the gauge coupling constants
in eq.\couplstr\
with the string parameters (see eqs.\gauge\ and \elim) one needs
an internal Calabi-Yau space with volume being slightly larger
that the inverse string scale $M_s^{-1}$.
Concretely,
for the interesting case, when the string scale $M_s$ is 
chosen to 
be the GUT scale of $2\cdot 10^{16}~{\rm GeV}$
in accordance with \couplstr, we derive the estimate
\eqn\estimate{
M_s^6\ \fc{V_6}{(2\pi)^6}=\fc{1}{8}\ g_{st}^2\ \fc{M_{pl}^2}{M_s^2}\sim 
4.5\cdot 10^4\ g_{st}^2}
on the volume $V_6$ in units of the string mass $M_s$.
Therefore for a uniform Calabi-Yau space, with 
\eqn\volsix{
V_6=(2\pi)^6\ R^6\ ,}
this condition \estimate\ boils down to:
\eqn\boil{
M_s^6\ R^6\sim 4.5\cdot 10^4\ g_{st}^2\ .}
Assuming $g_{st}=g_X$ this requirement is achieved for 
a uniform Calabi-Yau with radius $R$ being of the size
\eqn\eachrad{
M_s R=5.32\ .}
In addition,
the complex structure moduli of the Calabi-Yau  have to be tuned such that
the individual couplings at the string scale match the values shown
in eq.\couplstr. Via eq.\elim\ this requires the following values for
$V_a/\sqrt{V_6}$:
\eqn\vas{
V_w=2\ V_s=0.235\ \sqrt{V_6}\, .}
Setting $V_a=(2\pi)^3R_a^3$ and assuming again  $g_{st}=g_X$
this can be immediately translated to
\eqn\ras{M_s R_s=2.6, \quad M_s R_w=3.3   .}
Note that in our discussion we have neglected the one-loop threshold
corrections $\Delta_a$. The latter can be shown to be indeed small
for the ${\cal N}=1$ sectors of the toroidal model of \CremadesVA\ by using the explicit
one-loop threshold formula derived in \LustKY.

Let us compare the intersecting brane world picture
with the heterotic unification scenario. 
As emphasized already, at tree level the heterotic
string scale is independent from the internal Calabi-Yau volume
and is determined only
in terms of the Planck mass and the string coupling constant with the
effect that gauge coupling unification with the MSSM spectrum
is not possible.
One way to get consistent heterotic unification 
is to include the effect of one-loop gauge threshold corrections.
Specifically it was shown in \IbanezZV\
that minimal string unification 
in heterotic orbifolds with MSSM spectrum in addition to
 moduli dependent
threshold
corrections is in principle possible provided 
that the radius of the orbifold space is enlarged compared
to $M_s^{-1}$.
So assuming a non-perturbative S-duality among the heterotic
string and the brane world models, the splitting of the heterotic
gauge couplings at $M_s$ due
the heterotic one-loop threshold corrections $\Delta_a$
is mapped
to the various wrapped brane volumes $V_a$ at string tree level in the
brane world models.
Furthermore, the smallness of the gauge couplings
$\alpha_a(M_s)$ in heterotic string
compactifications is related to the universal
value of the heterotic string coupling $g_{st}$
plus the size of the $\Delta_a$; on the other hand in the dual
intersecting brane world models, the small values for $\alpha_a(M_s)$
are essentially due to the small ratio $M_s/M_{pl}$.

\vskip 0.3cm
\noindent
\vbox{
\hbox{\noindent\epsfysize=10truecm\epsfbox{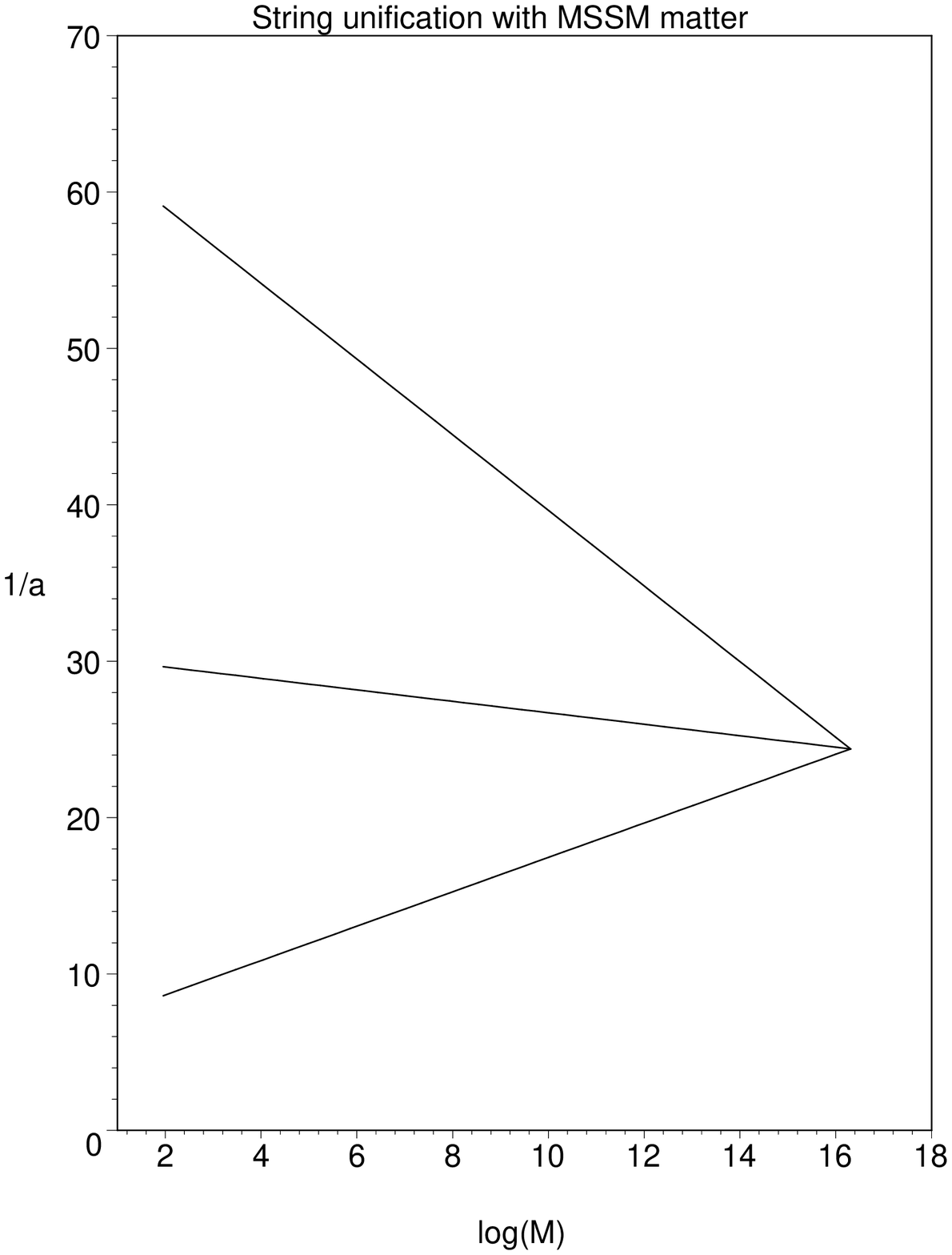}\hskip 0.5cm
\epsfysize=10truecm\epsfbox{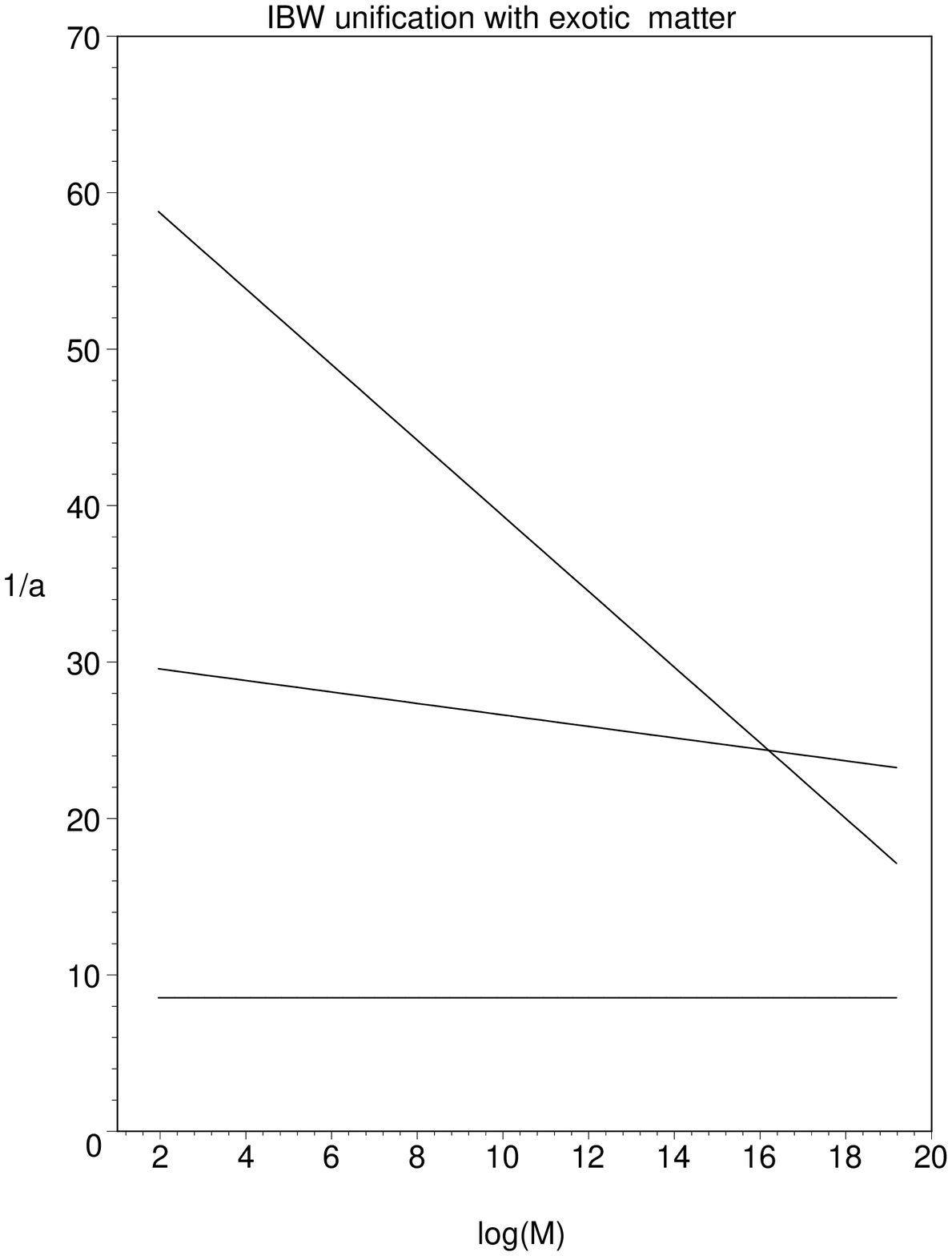} }
\noindent
\hbox{\hskip 3.2cm Fig.2 \hskip 5.8cm Fig.3 }
\bigskip\noindent
}
%\fig{}{plot1.eps}{8truecm}
%\noindent

\newsec{Exotic matter}

In general besides the chiral matter  string theory contains also additional
non-chiral matter. This is also localized on higher dimensional intersection
loci of the $D6$ branes and also comes with multiplicity $n_{ij}$ with
$i,j\in\{a,b,c,d\}$. Here $n_{ij}$ for $i\ne j$ counts the number of 
hypermultiplets in bifundamental respectively (anti-)symmetric 
representations of the gauge group, 
whereas for $i=j$ the number $n_{ii}$ denotes
the number of chiral multiplets in the adjoint representation
of the gauge group. Adding such exotic matter  to the MSSM matter
changes of course the individual beta-function coefficients, but for the 
combination $B$ one obtains
\eqn\bbb{ B={12}  - 2\, n_{aa} - {4}\, n_{ab} + 2\, n_{a'c} 
         + 2\, n_{a'd} - 2\, n_{bb}  +
             2\, n_{c'c} +  2\, n_{c'd}  + 2\, n_{d'd} }
which is always an even integer. Thus, even with exotic matter
the beta-function combination $B$ can only vary in steps of two. 
Note, that some sectors like $n_{bc}$  or $n_{a'a}$ have dropped out 
completely and 
that some representations contribute positively
and others negatively to $B$, so that cancellations are possible. 
In Table 3, utilizing the relation  \rela, we show how the string scale 
depends  on the beta-function  parameter $B$ 
\vskip 0.8cm 
\vbox{ \centerline{\vbox{
\hbox{\vbox{\offinterlineskip
\def\tablespace{height2pt&\omit&&\omit&&\omit&&\omit&&\omit&&\omit&& 
 \omit&\cr}
\def\tablerule{\tablespace\noalign{\hrule}\tablespace}

\hrule\halign{&\vrule#&\strut\hskip0.2cm \hfill #\hfill\hskip0.2cm\cr
& ${\bf B}$  && 18  && 16   && 14 && 12&& 10 && 8   &\cr
\tablerule
& ${\bf M_s}[{\rm GeV}]$ &&  $3.36\cdot 10^{11}$ &&$5.28\cdot 10^{12}$  && 
${ 1.82\cdot 10^{14}}$ && $2.04\cdot 10^{16}$ && $1.51\cdot 10^{19}$ &&
$3.06\cdot 10^{23}$  &\cr
}\hrule}}}}
\centerline{
\hbox{{\bf Table 3:}{\it ~~ $M_s(B)$ }}}
}
Interestingly for $B=10$ there exists a value for $M_s$ which is of the order
of the Planck scale
\eqn\order{       {M_{s}\over M_{pl}}=1.24\sim \sqrt{\pi\over 2} .}

One example of this type is  for instance given by
choosing $n_{aa}=1$, in which 
case the beta-function coefficients read $(b_3,b_2,b_1)=(0,-1,-11)$.
The couplings at the string  scale turn out to be
\eqn\couplstri{  \alpha_s(M_s)=0.117, \quad \alpha_w(M_s)=0.043, \quad
                  \alpha_Y(M_s)=0.035 }
leading to $\sin^2 \theta_w(M_s)=0.445$.
As can be seen from figure 3, in this case not all  three couplings 
intersect in one point, but nevertheless satisfy  the relation
\rela\ at the string  scale. 
For the scales  of the overall Calabi-Yau volume and the 3-cycles
we obtain
\eqn\exseize{   M_s R=0.6, \quad M_s R_s=1.9,\quad M_s R_w=3.3 .}

\newsec{Conclusions}

In this paper we have shown that under a few natural assumptions,
realistic three generation supersymmetric intersecting brane world models 
lead to gauge coupling unification.
We have argued that this result is quite robust against
stringy contributions of additional exotic non-chiral matter.
It is interesting to note that  using the MSSM spectrum 
the relation \hgaugeb\  
is accompanied by the accidental relation $\alpha_w=\alpha_s$. 
These relations are  compatible with an $SU(5)$ or $SO(10)$ GUT,
though emerge a priori without any reference to a simple gauge group.
%but rather is only due to the gauge group
%structure and 
%the restrictions on the intersection numbers we have imposed.
Therefore, one might wonder 
%It is however 
%not excluded
whether these relations find a group theoretical explanation by a 
(partial) gauge group unification which could happen
in some decompactification limit where more branes are lying on top
of each other.

So we like to emphasize again that the results presented in this article
are due to a bottom up approach where we start from a few well
motivated phenomenological assumptions. However to best of our knowledge
so far there exists no explicit intersecting brane world construction
which features all the necessary requirements, in particular
provides the spectrum of the 3 generation MSSM without any exotic
particles.
In view of this, the challenge is even more pressing 
to construct realistic supersymmetric intersecting brane world models
with a matter content satisfying $B\in\{10,12\}$. With such a model
one could compute more detailed informations about, for instance,
whether the
complex structure moduli can be chosen in such a way that eq.\vas\
is fulfilled, 
and about the 
string threshold corrections \doubref\TW\LustKY. 
We hope to report on this issue in the future
\rbgotwo. 

\vskip 1cm
\centerline{{\bf Acknowledgements}}\pano
This work  is supported in part by the EC
RTN programme HPRN-CT-2000-00131, the Deutsche Forschungsgemeinschaft (DFG), and 
the German--Israeli Foundation (GIF).
%\vfill\eject

\listrefs

\bye
\end